\documentstyle[12pt,epsfig]{article}
\begin{document}
\renewcommand{\floatpagefraction}{.80}
\renewcommand{\topfraction}{.90}
\renewcommand{\bottomfraction}{.90}
\title{ Solution of the Dyson equation for nucleons in the superfluid phase}
\author{
J.~Terasaki$^{a,b}$, 
F.~Barranco$^c$, 
R.~A.~Broglia$^{a,b,d}$, 
E.~Vigezzi$^{b}$ \\
and P.F.~Bortignon$^{a,b}$
\vspace{10pt}  \\
$^a$ Department of Physics, University of Milan,
Via Celoria 16, \\
I-20133 Milan, Italy \\
$^b$ INFN Sezione di Milano,
Via Celoria 16, I-20133 Milan, Italy \\
$^c$ Departamento de F\` \i sica Aplicada III, \\ 
Escuela Superior de Ingenieros,\\
Universidad de Sevilla, Camino de los Descubrimientos, \\
Sevilla, Spain\\
$^d$ The Niels Bohr Institute, University of Copenhagen, \\
DK-2100 Copenhagen, Denmark \\
}
% end of author
%
\date{May 23, 2001}
\maketitle
\begin{abstract}
We investigate the role the interweaving of surface vibrations 
and nucleon motion  has on Cooper pair formation in spherical 
superfluid nuclei.   
A quantitative calculation of the
state-dependent pairing gap requires to go beyond the quasiparticle
approximation, treating in detail the breaking of the single-particle
strength and of the associated poles.
This is done solving self-consistently the Dyson equation, 
including both
a bare nucleon-nucleon interaction (which for simplicity we choose 
as a monopole-pairing force of constant matrix elements $g$) and 
an induced interaction arizing from the exchange of vibrations 
(calculated microscopically in QRPA) between pairs of nucleons 
moving in time reversal states. 
Both the normal and anomalous density Green functions are included, thus
treating self-energy and pairing processes on equal footing.
We apply the formalism to the superfluid nucleus $^{120}$Sn. 
Adjusting the value of $g$ so as to reproduce, for levels close to the 
Fermi level, the empirical odd-even mass difference
($\Delta$ $\approx$ 1.4 MeV), it is found that the pairing gap  
receives about equal contributions from the monopole-pairing force and 
from the induced interaction. 
This result is also reflected in the fact that 
if one were to reproduce the observed $\Delta$  allowing the nucleons 
to interact only through the bare monopole-pairing force, 
a value of $g \approx 0.233$ MeV ($\approx 28/A$ MeV) is needed, 
50\% larger than the value $g \approx 0.166$ MeV  ($\approx 20/A$ MeV) 
needed in the full calculation. 
To keep in mind that the bare and the induced pairing contributions 
to $\Delta$  enter the corresponding equations in a very 
nonlinear fashion.  
It is furthermore found that self-energy processes reduce the 
contribution of the phonon induced interaction to the pairing gap 
by $\approx 20\%$ as compared to the value obtained by 
only phonon exchange without taking into account the 
breaking of the single-particle strength. 
\end{abstract}
 
 

\section{Introduction}

It is well known that single-particle motion in atomic nuclei 
is strongly renormalized
by the coupling to low-lying surface vibrations, which leads to the 
fragmentation of the single-particle strength and affects basic quantities 
like the level density at the Fermi surface, the effective mass and 
the width of giant resonances [1--3].
Recently, 
it has been shown that the induced interaction resulting from such 
a coupling (phonon-induced interaction) leads to pairing gaps 
which account for about half of the experimental values [4]. In what follows
we shall refer to these calculations of the state-dependent pairing gap
as the Bardeen-Cooper-Schrieffer plus Bloch-Horowitz (BCS+BH) approximation,
in keeping with the fact that this form of perturbation theory was used to
calculate the induced   interaction entering the gap equation and 
arising from the exchange of phonons between pairs of
nucleons moving in time reversal states close to the Fermi level.
In these calculations, empirical single-particle levels were
used, and the interplay between core polarization 
and self-energy contributions,
as well as  
the energy-dependence of the effective mass [2] was 
neglected.

In the present paper we take a step towards a more consistent
calculation of the renormalization effects associated with the 
particle-vibration coupling phenomenon, treating self-energy and
induced pairing interaction processes on equal footing (cf. Fig.\ 1(a)),
by solving the Dyson equation (also called Nambu-Gorkov equation, 
cf. e.g. [5]),
describing the motion of nucleons and their contribution to both normal
and anomalous densities.
Moreover and following ref.\ \cite{Ba99}, we add to this interaction 
a monopole-pairing force acting on the orbitals close to the Fermi level.

\begin{figure}
\begin{center}
\epsfig{file=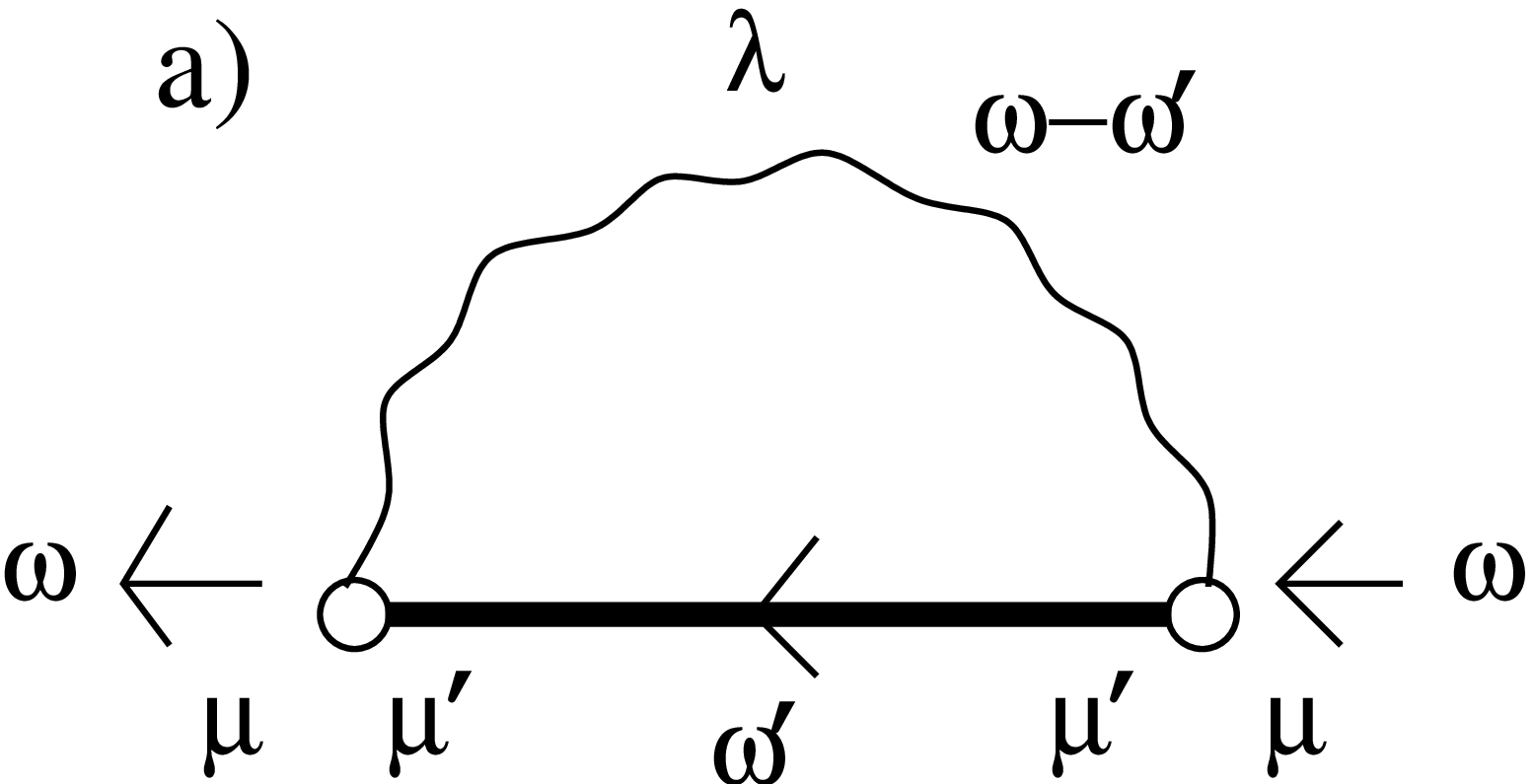,height=3cm}

\vspace{2em}

\epsfig{file=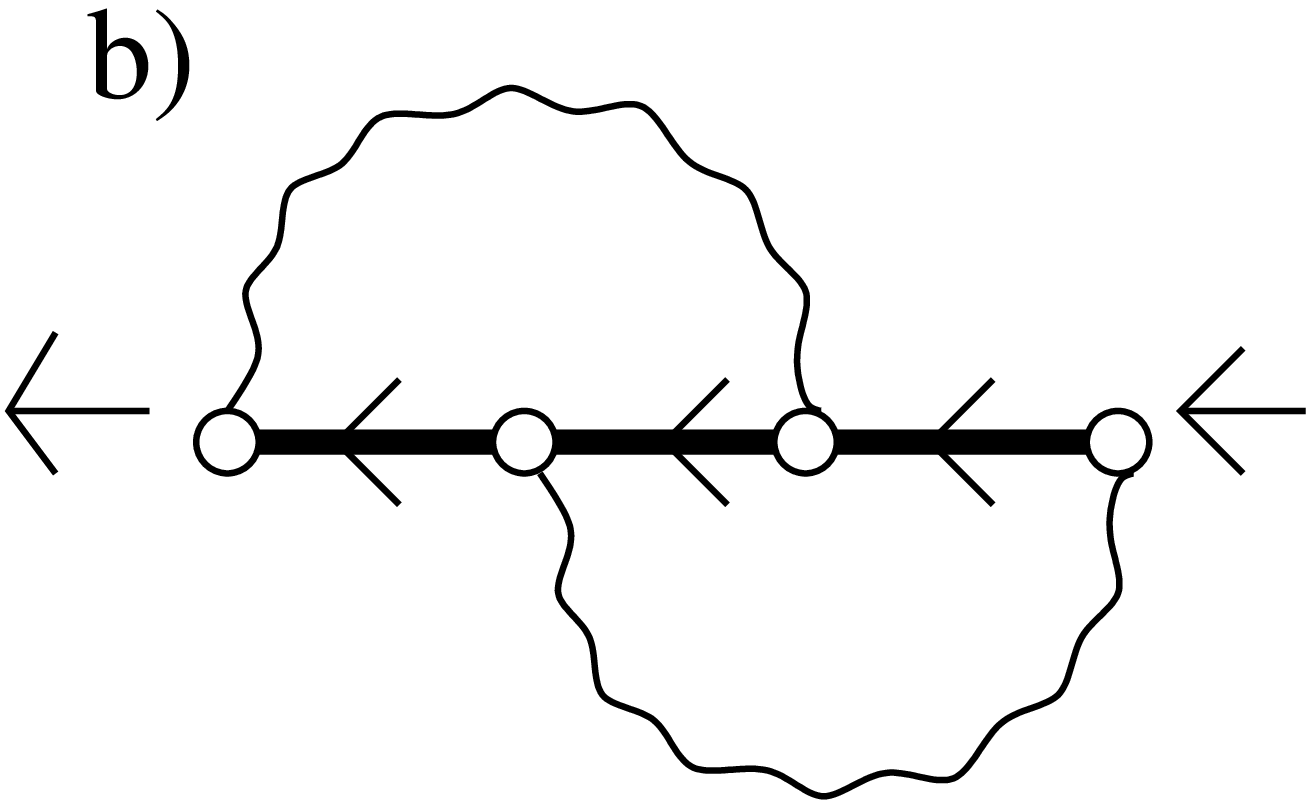,height=3cm}
\end{center}
\begin{center}
% osfmi:/temp/terasaki/tex/1stpaper/fig1b.gnu
% osfmi:/temp/terasaki/tex/1stpaper/fig1a.gnu
\parbox{12cm}{\small {\bf Figure 1.}
a) The self-energy $\hbar\Sigma_\mu^{\rm pho}(\omega)$ used
in the calculations. 
The wavy line denotes the unperturbed phonon Green function. 
The heavy arrowed line indicates the perturbed nucleon Green function. 
The small open circle is the vertex of the particle-phonon 
interaction. 
b) Example of self-energy diagram which includes the vertex correction.
}
\end{center}
\end{figure}

No vertex corrections  (cf. e.g. graph (b) of Fig.\ 1 )
have been considered in the present calculations,
because in the case of $^{120}$Sn, the nucleus we use in
the present paper to exemplify the theory,
vertex correction
contributions are found to change (reduce) the value of the state-dependent
pairing gap around Fermi level only modestly.\footnote{
One could argue concerning the validity of the approximation 
to neglect the vertex correction in terms of 
Migdal theorem [6,7] which states that the contribution of the vertex
renormalization graphs to the total electron-phonon coupling 
strength is smaller than the lowest-order strength by a
factor $\hbar \Omega/E_F$, 
where $\hbar\Omega$ is a typical phonon energy, $E_F$ being 
the Fermi energy measured from the bottom of the single-particle potential.  
Because in the nuclear case $E_F \approx $ 36 MeV and 
$\hbar \Omega$ is of the order of 
few MeV, the ratio above turns out to be approximately 
equal to $10^{-1}$. However, this estimate can hardly be considered 
quantitative, because of the marked shell structure displayed
by the single-particle levels,
and by the associated particle-vibration coupling matrix elements
(spin-flip vs. non spin-flip processes) as well as 
the strong cancellation existing between
effective mass and vertex correction contributions associated with
core polarization phenomena (cf. e.g. [3]).}

The Dyson equation  approach has been customarily used in condensed matter
physics to deal with strong coupling superconductors (cf. e.g. ref.\ [5--9]
and refs. therein).  In nuclear physics, this scheme has also
been used previously both in the case of nuclear matter  and of finite
nuclei [10--15], the work of ref.\ [15] being close in spirit to the one
discussed here. 
However in this reference the focus was set on applying the method to correct
the phenomenological values of single-particle  energies and of the pairing
gap to avoid double counting arising in connection with particle-vibration
coupling processes.
Furthermore, since we do not make use of the 
linear approximation adopted in that reference to solve the Dyson equation,
the present approach
should be also able to deal with strong-coupling situations.

 In section 2 we present the formalism used in solving the Dyson equation. 
In section 3 we discuss the results obtained  including
only the phonon-induced interaction, while in
section 4 we add the monopole-pairing force to it. Conclusions are
collected in section 5.

\section{Solution of the Dyson equation}

The core of the present  study is the Dyson equation, which for paired 
systems  is written in a matrix form, according to Nambu-Gorkov theory [5]: 
\begin{eqnarray}
 &&G^{-1}_\mu (\omega) = {G^0_\mu}^{-1} (\omega) - \Sigma^{\rm pho}_\mu(\omega) \ ,
 \label{Dy} \\
&&G_\mu(\omega) =
\left(
 \begin{array}{cc}
  G_\mu^{11}(\omega) & G_\mu^{12}(\omega) \vspace{5pt} \\
  G_\mu^{21}(\omega) & G_\mu^{22}(\omega) 
 \end{array}
\right) \ ,
%\\
%
%&&G_\mu^0(\omega) =
%\left(
% \begin{array}{cc}
%  {G_\mu^0}^{11}(\omega) & {G_\mu^0}^{12}(\omega) \vspace{5pt} \\
%  {G_\mu^0}^{21}(\omega) & {G_\mu^0}^{22}(\omega) 
% \end{array}
%\right) \ ,
\end{eqnarray}
$\omega$ being the energy variable. 
$G_\mu(\omega)$ denotes the perturbed single-particle (quasiparticle)
Green function, and its major poles  
%calculated self-consistently by iteration, taking also into account 
%the
%strength (residue at the pole) of the single-particle, 
yield the  quasiparticle energies of the system.
The diagonal and off-diagonal elements of Eq.\ (2) are the particle-hole and 
the pairing Green function respectively.
$G_\mu^0(\omega)$ is the unperturbed single-particle Green function,
with elements given by,
\begin{eqnarray}
&&\frac{1}{\hbar}{G_\mu^0}^{11}(\omega) =
\frac{e^{i\eta\omega}}
     { \omega - ( \varepsilon_\mu^0 - \varepsilon_F) + i\eta_\mu } \ , \\
&& \frac{1}{\hbar}{G_\mu^0}^{12}(\omega) =
   \frac{1}{\hbar}{G_\mu^0}^{21}(\omega) = 0 \ , \\
&& \frac{1}{\hbar}{G_\mu^0}^{22}(\omega) =
\frac{e^{-i\eta\omega}}
     { \omega + ( \varepsilon_\mu^0 - \varepsilon_F) - i\eta_\mu } \ , 
\end{eqnarray}
where $\varepsilon_\mu^0$ denotes the 
unperturbed single-particle energy of the single-particle state with 
quantum numbers 
$\mu$ ( $\equiv nlj$). 
The Fermi level  measured from the top of the potential is indicated by
$\varepsilon_F $. The parameter $\eta_\mu$ is defined as
\begin{equation}
\eta_\mu = 
\left\{
 \begin{array}{cc}
  -\eta \ , & \varepsilon_\mu^0 < \varepsilon_{\rm F}\ , \\
   \eta \ , & \varepsilon_\mu^0 > \varepsilon_{\rm F}\ ,
 \end{array}
\right.
\end{equation}
where $\eta$ is real and  positive.
The self-energy term of the phonon-induced interaction (cf. Fig.\ 1(a))
is defined as \cite{Sc64},
\begin{eqnarray}
\hbar\Sigma^{\rm pho}_\mu(\omega)
 &=&
\hbar
\left(
 \begin{array}{cc}
  \Sigma_\mu^{11}{}^{\rm pho}(\omega) & \Sigma_\mu^{12}{}^{\rm pho}(\omega)
  \vspace{5pt} \\
  \Sigma_\mu^{21}{}^{\rm pho}(\omega) & \Sigma_\mu^{22}{}^{\rm pho}(\omega) 
 \end{array}
\right) \nonumber \\
&=& 
 \int_{-\infty}^\infty \frac{d\omega^\prime}{2\pi} 
 \sum_{\mu^\prime}
 \tau_3
 \frac{1}{\hbar} G_{\mu^\prime}(\omega^\prime)
 \tau_3
 \sum_{\lambda n} 
 \frac{\hbar}{ 2\Omega_{\lambda n} B_{\lambda n} }
 \frac{1}{2j_\mu+1} \nonumber \\
&&
 \times \left|
 \langle \mu || R_0 \frac{\partial U}{\partial r} Y_\lambda ||\mu^\prime
 \rangle \right|^2
 \frac{i}{\hbar} D_{\lambda n}^0 (\omega-\omega^\prime) \ , \label{s-eint} 
\end{eqnarray}
where
$
\tau_3 = 
\left(
{\scriptsize
\begin{array}{rr}
\small
\small
1 & 0 \\
0 & -1 
\end{array}
}
\right). 
$
The energy and the inertial parameter of the phonons are denoted by
$\hbar \Omega_{\lambda n}$ and by $B_{\lambda n}$, respectively, 
with the multipolarity $\lambda$ and an additional index $n$.    
The particle-vibration vertex is given by 
\begin{equation}
 (-)^{j_\mu + j_{\mu^\prime}}
 \sqrt{\frac{\hbar}{ 2\Omega_{\lambda n} B_{\lambda n}} }
 \langle \mu || R_0 \frac{\partial U}{\partial r} Y_\lambda ||\mu^\prime
 \rangle \ ,
\label{vertex}
\end{equation}
where $R_0$ and $U(r)$ are the nuclear radius and potential.
The phonons
are not dressed and are calculated within the  QRPA,
as was done in ref.\ \cite{Ba99}. 
The corresponding (unperturbed) phonon propagator is
$$
 \frac{i}{\hbar} D_{\lambda n}^0 (\omega-\omega^\prime) = 
 \frac{i}{ \omega-\omega^\prime -\hbar\Omega_{\lambda n} + i\eta_D }
 + 
 \frac{i}{-\omega+\omega^\prime -\hbar\Omega_{\lambda n} + i\eta_D } \ ,
%\label{phonon}
$$
where $\eta_D$ stands for a real positive parameter. 

 We solve Eqs.\ (1) and (7) self-consistently,  starting 
by inserting  at  the place of $G_{\mu'}$ in Eq.\ (7)
the BCS Green function in an analytical form 
(see Chap.\ 7 in \cite{Sc64}), calculating the integral of the self-energy
$\Sigma^{\rm pho}_{\mu}(\omega)$ analytically.
The poles $\pm \omega_{G+}^{\mu m}$ 
(we set ${\rm Re}\,\omega_{G+}^{\mu m} \ge 0$, 
$m$ being another label distinguishing the pairs of the poles )
of the new Green function 
are determined by  searching, on the real axis, 
for the roots of the equation\footnote{We use a mesh of 1 keV,
and once the existence of a root is detected 
in an interval, the bisection method is applied. }
\begin{equation}
\det \overline{ G}_\mu^{-1} (\pm {\rm Re}\,\omega_{G+}^{\mu m}) = 0 \ ,
\label{eqdet}
\end{equation}
where 
$\overline{ G}_\mu^{-1} (\omega)$ 
is calculated from Eq.\ (\ref{Dy}) but neglecting 
the 
non-hermitian components:
\begin{equation}
\hbar\overline{G}_\mu^{-1} (\omega) = 
\left(
 \begin{array}{cc}
  \omega -\tilde{\varepsilon}_\mu^0
  -{\rm Re}\,\hbar\Sigma_\mu^{11}{}^{\rm pho}(\omega) &
  -\hbar\Sigma^{12}_\mu{}^{\rm pho}(\omega) \vspace{10pt} \\
  -\hbar\Sigma^{21}_\mu{}^{\rm pho}(\omega) &
  \omega +\tilde{\varepsilon}_\mu^0
  -{\rm Re}\,\hbar\Sigma_\mu^{22}{}^{\rm pho}(\omega)
 \end{array}
\right) \ ,
\label{Gbarinv}
\end{equation}
with $\tilde{\varepsilon}_\mu^0 = \varepsilon_\mu^0 - \varepsilon_F$. 
A very small value (1 keV) of the averaging parameters 
$\eta $ and $\eta_D$ is used with an idea to deal with stationary states.
The imaginary parameters are so small that we can identify
the residue of $\overline{G}_\mu^{11}(\omega)/\hbar$ with 
that of $G_\mu^{11}(\omega)/\hbar$, that is, 
\begin{equation}
R^{11}_{\mu m}(\pm \omega_{G+}^{\mu m}) \simeq
\left(
 \left.
 \frac{d}{d\omega}
 \frac{\hbar}{\overline{G}_\mu^{11}(\omega)}
 \right|_{\pm {\rm Re}\,\omega_{G+}^{\mu m}}
\right)^{-1} \ ,  \label{Z11}
\end{equation}
where the derivative is taken also on the real axis. 
$R^{12}_{\mu m}(\pm\omega_{G+}^{\mu m})$ is obtained through 
a similar equation.
%
%For the pairing part the residue is calculated
%%
%\begin{equation}
%Z^{12}_{\mu m}(\omega_{G+}^{\mu m}) = 
% \left.
% \frac{ - \Sigma_\mu^{12}(\omega) }
%  { \frac{d}{d\omega} {\rm det}\,{\overline{G}}^{-1}_\mu(\omega) }
% \right|_{ \omega={\rm Re}\,\omega_{G+}^{\mu m}}
%             \ .  \label{Z12}
%%
%\end{equation}
%
The imaginary part of the poles 
$\pm \omega_{G+}^{\mu m} $ is fixed to be 
$\mp$ 1 keV 
for simplicity. 
%
%\begin{equation}
%{\rm Im} \,\omega_{G+}^{\mu m} \simeq
%{\rm Im}\, \frac{\hbar}{G_\mu^{11}({\rm Re}\,\omega_{G+}^{\mu m})}
%R^{11}_{\mu m}(\omega_{G+}^{\mu m}) \ . \label{Impole}
%\end{equation}
%
Eqs.~(\ref{eqdet})--(\ref{Z11}) are generalizations 
of those found in Sections 
 3.4 and 4.3.6 of ref.\ \cite{Ma85}.
The new Green function can now be written as
\begin{eqnarray}
&&
\frac{1}{\hbar}G_\mu^{11}(\omega) =
\sum_m
 \left(
  \frac{R_{\mu m}^{11}( \omega_{G+}^{\mu m})}{\omega - \omega_{G+}^{\mu m} }
 +\frac{R_{\mu m}^{11}(-\omega_{G+}^{\mu m})}{\omega + \omega_{G+}^{\mu m} }
 \right) e^{i\omega\eta} \ ,
 \label{G11} \\
&&
\frac{1}{\hbar}G_\mu^{12}(\omega) =
\sum_m
\left(
  \frac{R_{\mu m}^{12}( \omega_{G+}^{\mu m})}{\omega - \omega_{G+}^{\mu m} }
 +\frac{R_{\mu m}^{12}(-\omega_{G+}^{\mu m})}{\omega + \omega_{G+}^{\mu m} }
\right)
\ . \ 
 \label{G12}
\end{eqnarray}
Assuming that the ground state of the system is time reversal invariant,
the quantities
$G_\mu^{22}(\omega)$ and $G_\mu^{21}(\omega)$ can be written 
in a similar way in terms of 
$R_{\mu m}^{11}(\pm\omega_{G+}^{\mu m})$ and 
${R_{\mu m}^{12}}^\ast(\pm\omega_{G+}^{\mu m})$, respectively.
A new self-energy is then generated, using Eqs.\ (12) and (13), as
\begin{eqnarray}
\hbar\Sigma_\mu^{11}{}^{\rm pho}(\omega) &=&
 \sum_{\mu^\prime m} \sum_{\lambda n} 
 \frac{\hbar}{ 2\Omega_{\lambda n} B_{\lambda n} } 
 \left|
 \langle \mu ||
 R_0 \frac{\partial U}{\partial r} Y_\lambda
 || \mu^\prime \rangle \right|^2 
 \frac{1}{2j_\mu+1} \:
 \nonumber \\
&&
 \times
 \left\{
  \frac{R^{11}_{\mu^\prime m}(-\omega_{G+}^{\mu^\prime m})}
   {\omega + \omega_{G+}^{\mu^\prime m} + \hbar\Omega_{\lambda n} - i \eta_D}
  +
  \frac{R^{11}_{\mu^\prime m}( \omega_{G+}^{\mu^\prime m})}
   {\omega - \omega_{G+}^{\mu^\prime m} - \hbar\Omega_{\lambda n}+ i \eta_D}
 \right\} \ . \label{slf-e-f1}
\end{eqnarray}
The other elements can be obtained in a similar way, and
the  process is iterated until convergence.
The Fermi level  
$\varepsilon_F$ is fixed so as to obtain the  desired expectation 
value of the nucleon number 
\begin{equation}
\langle \hat{N} \rangle =
\sum_{\mu m} (2j_{\mu}+1)\,R^{11}_{\mu m}(-\omega_{\rm G+}^{\mu m})\ .
\label{num}
\end{equation}

 As can be seen From Fig.\ 1(a) and Eq.\ (\ref{Dy}), this iterative process 
involves the coupling 
to an increasingly larger number of phonons at each iteration step.
This in turn enhances 
the fragmentation of the single-particle strength, leading to an 
increasingly larger 
number of poles of the Green function until saturation.\footnote{ 
The number of poles saturates because $\eta$ and $\eta_D$ are finite, and 
we use the mesh method for finding the poles as explained before. 
That is, the poles with very small residue (strength) are neglected.}
For each of the poles $\pm\omega_{\rm G+}^{\mu m}$, 
the sum 
$R_{\mu m}^{11}(\omega_{\rm G+}^{\mu m})
+ R_{\mu m}^{11}(-\omega_{\rm G+}^{\mu m}) $ is 
interpreted as the single-particle strength of the pole 
( $R_{\mu m}^{11}(-\omega_{\rm G+}^{\mu m}) = 
   R_{\mu m}^{22}(\omega_{\rm G+}^{\mu m})$ ).
As a rule,  and depending on whether pairing correlations are important or
not, one or two poles close to $\varepsilon_F$ carry most of the 
single-particle strength in the case of orbitals close to the Fermi level. 
In particular, in the BCS approximation
the two poles  $\pm\omega_{\rm G+}$
carry the whole strength,
and $R_\mu^{11}(\omega_{\rm G+}^{\mu})
+ R_\mu^{11}(-\omega_{\rm G+}^{\mu}) = u_{\mu}^2 + v_{\mu}^2 =1$, 
and $|R_\mu^{12}(\pm\omega_{\rm G+}^{\mu})| = u_{\mu} v_{\mu}$.
 In general, however, 
one needs to take into account more poles, in order to exhaust
a substantial fraction of the sum rule\
\begin{equation}
\sum_{m} \Big(R^{11}_{\mu m}(\omega_{\rm G+}^{\mu m})
 + R^{11}_{\mu m}(-\omega_{\rm G+}^{\mu m})\Big) =1\ .
 \label{sumrule}
\end{equation}
We shall label with $ \mu m_0$
the pole carrying the major single-particle strength, 
and call it the quasiparticle pole.

\section{Induced interaction}

We have applied the  formalism described above to the calculation of the 
neutron Green functions of the nucleus 
$^{120}$Sn. In this section we shall concentrate on
the phonon-induced interaction, while in the next 
we shall add the monopole-pairing force with constant matrix elements 
as a model of the bare neucleon-nucleon force.

 The unperturbed single-particle basis $\varepsilon_\mu^0$ has been calculated
with a Woods-Saxon potential
using an effective mass $m_k = 0.7 m$, $m$ being the bare nucleon mass. 
With this value, self-energy effects are expected to lead to a sensible 
single-particle density close to the Fermi level 
(cf. also Section 4.6.3 in \cite{Ma85}). 
We have included in the calculation all the single-particle bound levels. 

\begin{table}
\begin{center}
% See osfmi:/temp/terasaki/vertex/eff/conv_coupl .
\parbox{12cm}{\small {\bf Table 1.}
The energies of the phonon modes $\hbar\Omega^{\rm eff}_{\lambda n}$
and coupling strength 
$\beta^{\rm eff}_{\lambda n}/\sqrt{2\lambda+1}$. 
Tables a), b), c) and d) are for $\lambda^\pi = 2^+$, 3$^-$, 4$^+$ and 
5$^-$, respectively. The lowest-energy modes ($n = 1$) were taken from a QRPA
calculation directly. 
The coupling strengths as well as the energies of the other modes 
were determined by the procedure shown in Eqs.\ (\ref{veff1})--(18).
}
\end{center}

\begin{center}
\small
\begin{tabular}{lll}
\multicolumn{3}{l}{a)}\\
\noalign{\vspace{1ex}}
\hline
\noalign{\vspace{1ex}}
\multicolumn{3}{c}{$\lambda^\pi=2^+$}\\
\noalign{\vspace{1ex}}
$n$ & $\hbar\Omega^{\rm eff}_{\lambda n}$ &
$\beta^{\rm eff}_\lambda/\sqrt{2\lambda+1}$ \\
    & [MeV] & \\
\noalign{\vspace{1ex}}
\hline
\noalign{\vspace{1ex}}
1   & \hspace{1ex}1.173 & \hspace{1ex}0.0554 \\
2   & \hspace{1ex}5.2   & \hspace{1ex}0.0134 \\
3   &            12.5   & \hspace{1ex}0.0447 \\
\noalign{\vspace{1ex}}
\hline
\noalign{\vspace{1.2em}}
\end{tabular}
\hspace{2em}
\begin{tabular}{lll}
\multicolumn{3}{l}{b)}\\
\noalign{\vspace{1ex}}
\hline
\noalign{\vspace{1ex}}
\multicolumn{3}{c}{$\lambda^\pi=3^-$}\\
\noalign{\vspace{1ex}}
$n$ & $\hbar\Omega^{\rm eff}_{\lambda n}$ &
$\beta^{\rm eff}_\lambda/\sqrt{2\lambda+1}$ \\
    & [MeV] & \\
\noalign{\vspace{1ex}}
\hline
\noalign{\vspace{1ex}}
1   & \hspace{1ex}2.423 & \hspace{1ex}0.0591 \\
2   & \hspace{1ex}5.57  & \hspace{1ex}0.0317 \\
3   &            10.0   & \hspace{1ex}0.0238 \\
4   &            21.0   & \hspace{1ex}0.0291 \\
\noalign{\vspace{1ex}}
\hline
\end{tabular}
\par
\mbox{}\par
\mbox{}\par
\begin{tabular}{lll}
\multicolumn{3}{l}{c)}\\
\noalign{\vspace{1ex}}
\hline
\noalign{\vspace{1ex}}
\multicolumn{3}{c}{$\lambda^\pi=4^+$}\\
\noalign{\vspace{1ex}}
$n$ & $\hbar\Omega^{\rm eff}_{\lambda n}$ &
$\beta^{\rm eff}_\lambda/\sqrt{2\lambda+1}$ \\
    & [MeV] & \\
\noalign{\vspace{1ex}}
\hline
\noalign{\vspace{1ex}}
1   & \hspace{1ex}2.470 & \hspace{1ex}0.0248  \\
2   & \hspace{1ex}8.0   & \hspace{1ex}0.0300  \\
3   &            12.0   & \hspace{1ex}0.0300  \\
4   &            15.0   & \hspace{1ex}0.0270  \\
\noalign{\vspace{1ex}}
\hline
\end{tabular}
\hspace{2em}
\begin{tabular}{lll}
\multicolumn{3}{l}{d)}\\
\noalign{\vspace{1ex}}
\hline
\noalign{\vspace{1ex}}
\multicolumn{3}{c}{$\lambda^\pi=5^-$}\\
\noalign{\vspace{1ex}}
$n$ & $\hbar\Omega^{\rm eff}_{\lambda n}$ &
$\beta^{\rm eff}_\lambda/\sqrt{2\lambda+1}$ \\
    & [MeV] & \\
\noalign{\vspace{1ex}}
\hline
\noalign{\vspace{1ex}}
1   & \hspace{1ex}2.402 & \hspace{1ex}0.0250  \\
2   & \hspace{1ex}8.0   & \hspace{1ex}0.0365  \\
3   &            13.0   & \hspace{1ex}0.0166  \\
4   &            21.0   & \hspace{1ex}0.0232  \\
\noalign{\vspace{1ex}}
\hline
\end{tabular}
\end{center}
\end{table}

 The computation time needed for our calculations depends strongly on 
the number of phonon modes $\lambda n$ included.
The full QRPA response for the multipolarities
$\lambda^{\pi} =2^+,3^-,4^+,5^-$ in the energy interval 
0--20 MeV used in ref.\ [4] consists of about two hundreds
phonon modes of energies $\hbar\Omega_{\lambda n }$ and zero-point amplitudes 
$\beta_{\lambda n}$ for each multipolarity. 
We include the four lowest phonons, one for each multipolarity, 
which give the largest contributions to the induced phonon interaction. 
We account for the effects of the other roots including only a few 
effective phonons of energy $\hbar\Omega^{\rm eff}_{\lambda n}$, distributed 
in the interval 0--20 MeV, choosing their effective strength 
so that when they are used in the calculation of ref.\ [4] 
they reproduce the state-dependent gap obtained there.
This is obtained, considering that the sum 
of the (asymmetrized) matrix elements of the induced interaction between 
two pairs $(j_{\nu})^2_{J=0},(j_{\nu'})^2_{J=0} $ due to the phonons
lying in an energy interval $[\Omega_a,\Omega_b]$, 
calculated according to the BH formalism, is given by 
\begin{equation}
v_{\nu\nu^\prime} = 
\frac { |\langle \nu'|| R_0 
{\frac{\partial U}{\partial r}} || \nu \rangle|^2 }
{(2j_\nu +1)(2j_{\nu'}+1)(2\lambda+1)}
\sum_{\Omega_{\lambda n}= \Omega_a}^{\Omega_b} 
{\frac {4\beta^2_{\lambda n}}
{E_0 - (e_{\nu} + e_{\nu'} + \hbar \Omega_{\lambda n})}}\ ,
\label{veff1}
\end{equation}
where $E_0$ and $e_{\nu }$ are the correlation energy of the ground state
and the absolute value of the single-particle energy 
with respect to the Fermi level, respectively. 
The effective strength of the phonon  
representing this interval is then chosen so as to satisfy the equations 
\renewcommand{\theequation}{\arabic{equation}a}
\begin{equation}
{\frac{(\beta^{\rm eff}_{\lambda n})^2}
{E_0 - (e_{\nu} + e_{\nu'} + \hbar \Omega^{\rm eff}_{\lambda n})}}= 
\sum_{\Omega_{\lambda n}= \Omega_a}^{\Omega_b} 
{\frac{\beta^2_{\lambda n}}
{E_0 - (e_{\nu} + e_{\nu'} + \hbar \Omega_{\lambda n})}}\ ,
\label{veff2}
\end{equation}
\begin{equation}
\addtocounter{equation}{-1}
\renewcommand{\theequation}{\arabic{equation}b}
\hbar\Omega^{\rm eff}_{\lambda n} = \hbar\Omega_b\ , 
\label{veff2b}
\end{equation}
\renewcommand{\theequation}{\arabic{equation}}
for the pairs $(j_{\nu})^2_{J=0},(j_{\nu'})^2_{J=0}$ giving the 
largest contribution to the
pairing gap for the multipolarity $\lambda$. 
The energies and zero-point amplitudes devided by 
$\sqrt{2\lambda+1}$ of the  effective 
phonons are listed in Table 1. 
Then the coupling strength
$$
\frac{\hbar}{2\Omega^{\rm eff}_{\lambda n}B^{\rm eff}_{\lambda n}} = 
\frac{(\beta^{\rm eff}_{\lambda n})^2}{2\lambda+1}\ ,
$$
is used for Eq.\ (\ref{slf-e-f1}).  
The BCS+BH calculation performed with this restricted 
ensemble of phonons reproduces the state-dependent pairing 
gaps of ref.\ [4] within a few per cent.   

 Because the number of the poles can vary from one iteration step to the next,  
the solution displays (small) fluctuations.
We obtained a reasonable  accuracy in the particle number 
$\langle {\hat N} \rangle = 70 \pm 0.1$ and 
negligible fluctuations in the perturbed single-particle levels 
and the state-dependent pairing gaps (definitions are given later)
for most of the orbitals.
The sum of the single-particle strength for a given orbital $\mu$ is 
as a rule larger than 0.90 in the valence shell.\footnote{
To be noted that of all the (hundreds or eventually thousands)
poles, only a fraction $(\leq 10-20$) carry an appreciable single-particle
strength for orbitals near the Fermi level.
Consequently, calculations made considering only these
poles explicitly lead to results which coincide within 10\%, with the
results of the more accurate calculation presented in this paper. 
}

\begin{figure}
\begin{center}
\epsfig{file=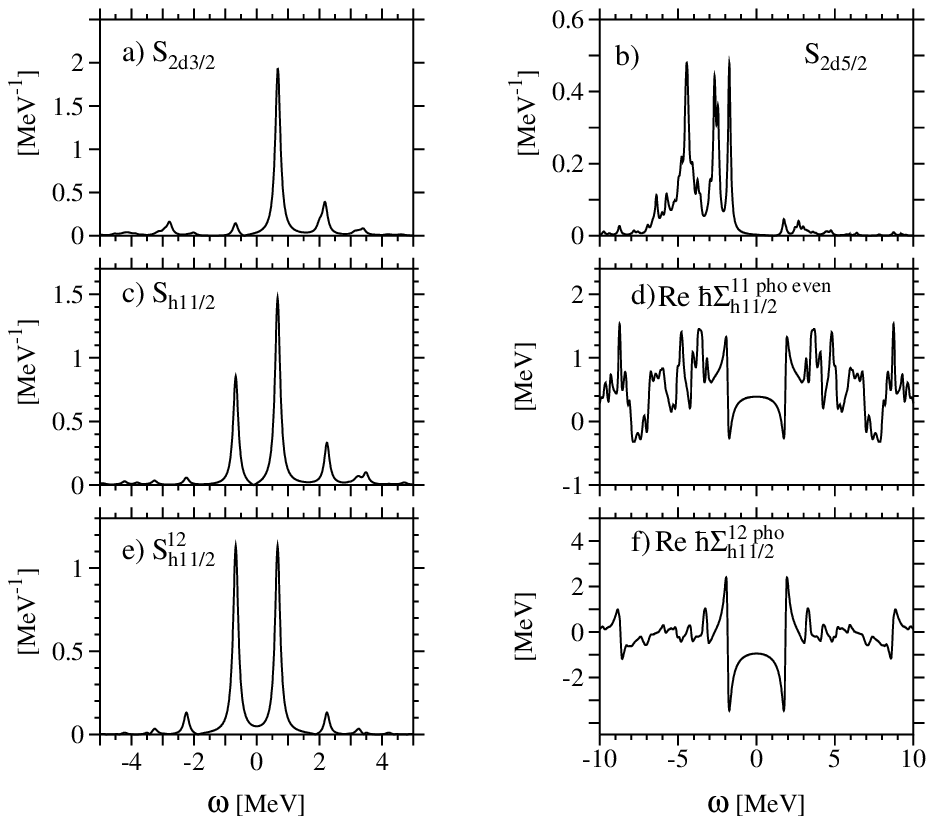,height=12cm}
\end{center}
\begin{center}
% pcprometeo:~/tmppool10/fig2.gnu <- old
% pcwalhalla:~/gs9/4-5pho/fig2.gnu
\parbox{12cm}{\small {\bf Figure 2.}
a) The spectral function of 2$d_{3/2}$  
calculated using the poles and residues of the solution 
of the Dyson equation. To display the results in terms of a 
smooth continuous curve, the results have been folded with a smooth function
with FWHM of 0.1 MeV. Similar foldings have been performed in connection
with the other plots.
b) Same but for the  2$d_{5/2}$ orbital ,
c) The spectral function of the 1$h_{11/2}$ orbital,
d) The even self-energy of the 1$h_{11/2}$ orbital 
(real part),
e) The pairing spectral function calculated in a similar way 
as the spectral function,
f) The pairing self-energy (real part). 
}
\end{center}
\end{figure}

The spectral function 
\begin{equation}
S_{\mu}(\omega) =
\frac{1}{\pi\hbar}|{\rm Im}\,G^{11}_{\mu} (\omega)|\ ,
\label{spectral}
\end{equation}
of some levels is shown in Fig.\ 2 (graphs (a),(b) and (c)). 
From this figure it can be seen that
the quasiparticle approximation
(keeping only one (no pairing) or two (with pairing) pole(s))
works well for the level $2d_{3/2}$, 
(particle-like quasiparticle,  cf. Fig.\ 2(a)), 
and 1$h_{11/2}$ (with an associated single-particle strength 
$R^{11}_{\mu m_0}(-\omega_{G+}^{\mu m_0}) \approx 0.3 $,
cf. Fig.\ 2(c)),
but breaks down 
for the  hole-like $2d_{5/2}$  level  (cf. Fig.\ 2(b)),
which displays a substantial degree of fragmentation,
because one of the poles of 2$d_{5/2}$ is almost degenerate with
the 1$h_{11/2} \times  3^-$ state,
then the single-particle strength (residue) of the pole becomes small 
(see Eqs.\ (\ref{slf-e-f1}), (\ref{Dy}) and (\ref{Z11})),
and the other poles carry more strength. 
In addition, a large matrix element connecting 
$1h_{11/2}$ and $2d_{5/2}$ with $\lambda^\pi\,n = 3^-\,1$ 
(see Eq.\ (\ref{vertex})) enhances the fragmentation. 
Besides the spectral function, we also display in Fig.\ 2
the even self-energy
\begin{equation}
\hbar{\Sigma_{\mu}^{11}{}^{\rm pho}}^{\rm even} (\omega)
= \frac{\hbar}{2} \Big(\Sigma_{\mu}^{11}{}^{\rm pho}(\omega) 
+\Sigma_{\mu}^{11}{}^{\rm pho}(-\omega)\Big)\ .
\label{even}
\end{equation}
This is shown in Fig.\ 2(d) for $\mu = 1h_{11/2}$.
Its value calculated at the quasiparticle peak of the spectral function gives
the approximate energy of the renormalized
single-particle spectra 
\begin{equation}
\varepsilon^1_{\mu} =
\frac{1}{Z_\mu(\omega_{\rm G+}^{\mu m_0})}
\Big(\tilde{\varepsilon}_{\mu}^0
+\hbar {\rm Re}\,{\Sigma^{11}_{\mu}{}^{\rm pho}}^{\rm even}
(\omega_{G+}^{\mu m_0})\Big)
+ \varepsilon_F\ ,
\label{singlemod}
\end{equation}
with
$$
Z_\mu(\omega_{\rm G+}^{\mu m}) =
1 - \frac{\hbar{\rm Re}\,{\Sigma_\mu^{11}{}^{\rm pho}}^{\rm odd}
(\omega_{\rm G+}^{\mu m})}
{{\rm Re}\,\omega_{\rm G+}^{\mu m}}\ .
$$ 
The function $\hbar{\Sigma_\mu^{11}{}^{\rm pho}}^{\rm odd}
(\omega_{\rm G+}^{\mu m})$ is the odd component of the self-energy. 
(See Chap.\ 7 in ref.\ \cite{Sc64} in connection to
$Z_\mu(\omega^{\mu m}_{G+}$).)  

\begin{figure}
\begin{center}
% pcwalhalla:~/gs9/4-5pho/new/fig3a.gnu
\epsfig{file=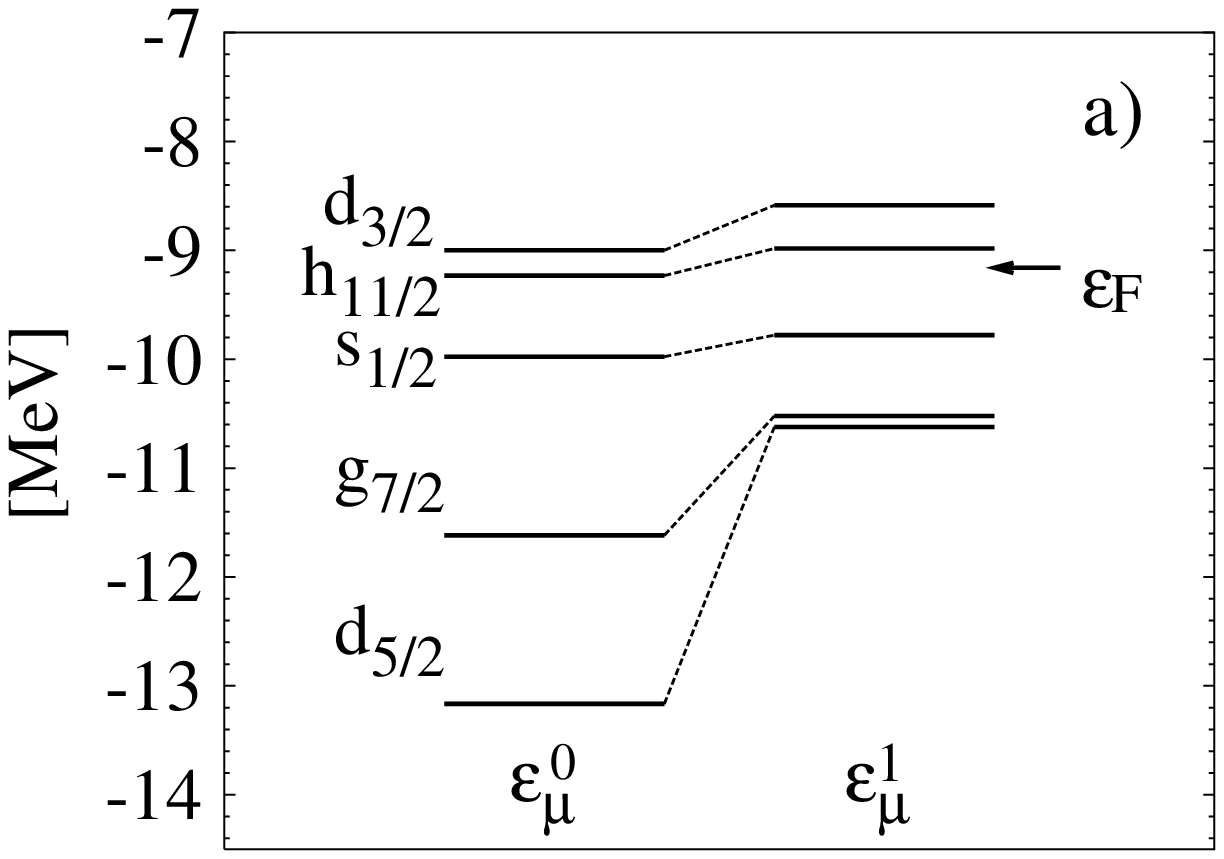,height=5cm}
\par\vspace{1em}
% pcwalhalla:~/gs9/4-5pho/new/fig3b.gnu
\epsfig{file=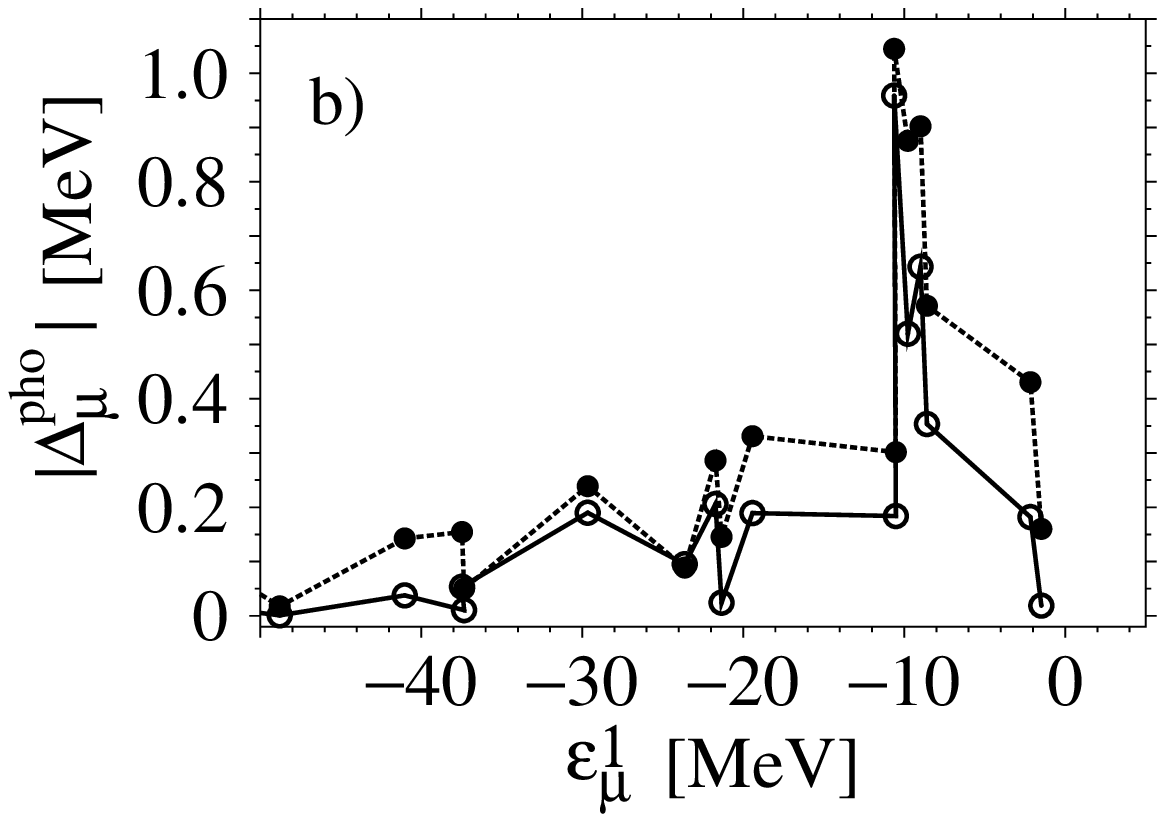,height=6cm}
\par\vspace{1em}
% pcwalhalla:~/gs9/4-5pho/new/fig3c.gnu
\epsfig{file=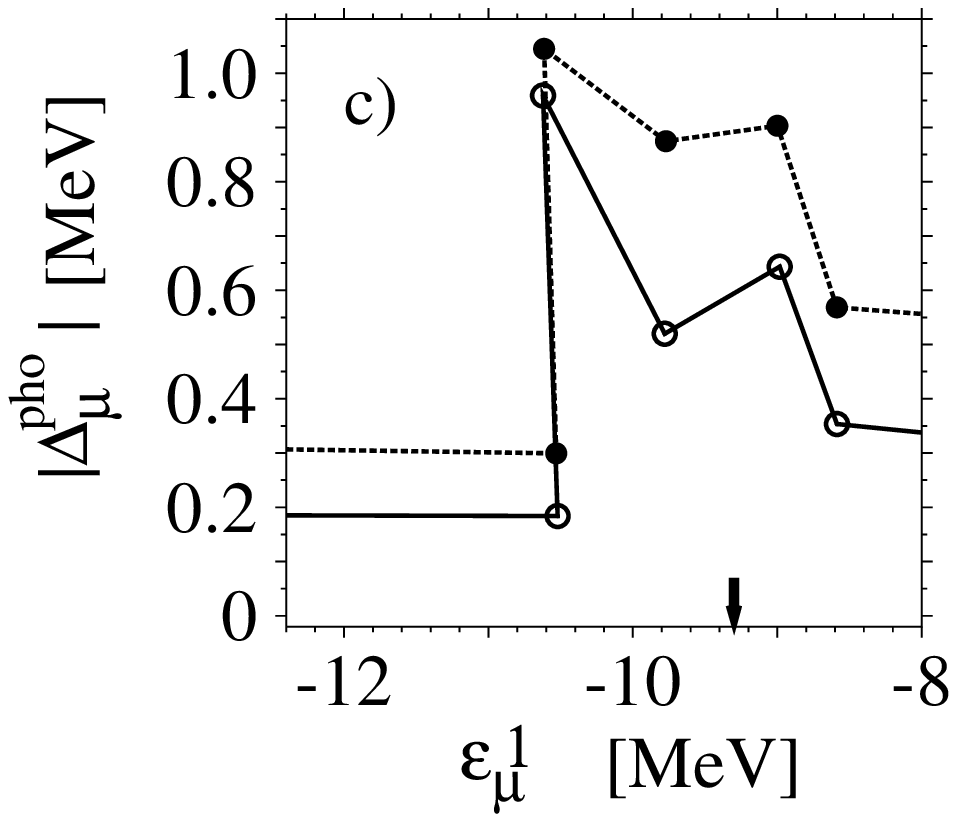,height=6cm}
\end{center}
\begin{center}
\parbox{12cm}{\small {\bf Figure 3.}
a) The unperturbed single-particle levels 
$\varepsilon^0_\mu$ near the Fermi level $\varepsilon_F$
and the perturbed levels $\varepsilon^1_\mu$ corresponding to the 
quasiparticle poles. 
b) The state-dependent pairing gaps corresponding to 
the quasiparticle poles (open circles) and 
the gaps of the BCS\-+BH calculation (filled circles) 
as a function of the renormalized (perturbed) single-particle energies
$\varepsilon_\mu^1$. 
c) A magnification of (b) near the Fermi level $\varepsilon_F$.  
The arrow indicates the Fermi level  associated with  the Dyson calculation. 
}
\end{center}
\end{figure}

The original  single-particle spectrum is 
compared in Fig.\ 3(a) with the energies 
$\varepsilon_\mu^1$ of the 
main (quasiparticle) peaks.
It is seen that the self-energy processes increase the level density in the 
valence shell.
This increase is essentially 
due to the rise of the energy of the hole
states. 

The spectral function and self-energy shown above have their
counterpart in the pairing channel. 
The pairing spectral function $S_\mu^{12}(\omega)$
corresponds to the anomalous density. In 
the BCS limit  it is 
peaked at the quasiparticle energy and corresponds to the product
$u_{\mu} v_{\mu}$.
The function 
$S_\mu^{12}(\omega)$ is shown in Fig.\ 2(e)
for $\mu = 1h_{11/2}$.
 The pairing self-energy $\hbar\Sigma^{12}_{\mu}{}^{\rm pho}(\omega)$
is shown in Fig.\ 2(f), 
again for the level $\mu =1h_{11/2}$. 
The complex state-dependent gaps are given by [5] 
\begin{equation}
\Delta_{\mu}^{\rm pho} =
\frac{\hbar\Sigma_\mu^{12}{}^{\rm pho}(\omega_{\rm G+}^{\mu m_0})}
{Z_\mu(\omega_{\rm G+}^{\mu m_0})}\ .
\label{complex}
\end{equation}
The assignment of a state-dependent pairing gap as well as 
a perturbed single-particle energy is valid only to the extent 
that the quasiparticle approximation is valid, 
thus the assignment may not be adequate for the single-particle orbital
$d_{5/2}$ (cf. Fig.\ 2(b)),
which is strongly fragmented. 
In this case, a tentative quasiparticle pole 
(the right major peak in Fig.\ 2(b)) was used. 
The results of the calculation are collected in Table 2, and 
the gaps associated to the quasiparticle peaks 
$\omega_{\rm G+}^{\mu m_0}$ are shown in Figs.\ 3(b) and (c)
in absolute value, where 
they are compared with the pairing gaps in the BCS+BH approximation 
calculated using Table 1 
and the renormalized (perturbed) single-particle spectra $\varepsilon^1_\mu$ 
of the Dyson calculation (these energies were used for calculating 
$e_\nu$ in Eqs.\ (\ref{veff1}) and (\ref{veff2}). 
Close to the Fermi level, see Fig.\ 3(c),  the gaps of Dyson calculation 
 are systematically lower;
differences of up to 400 keV between the two calculations are
observed.
In average in the valence shell, the difference is $\sim$ 200 keV. 

 It is also interesting to calculate more global quantities, 
like e.g. the pairing correlation energy.
Within the framework of the Dyson equation, the contribution to the total
ground state energy of the system arising from pairing correlations is 
\begin{eqnarray}
E^{\rm pho}_{\rm pair} &=& 
-i\sum_\mu (2j_\mu+1) \frac{1}{2} \int_{-\infty}^\infty 
\frac{d\omega}{2\pi}
\left(
\hbar\Sigma_\mu^{12}{}^{\rm pho}(\omega) \frac{1}{\hbar}
G_\mu^{21}(\omega) e^{i\eta\omega}
\right.
\nonumber\\
&&
\left.
+\ \hbar\Sigma_\mu^{21}{}^{\rm pho}(\omega) \frac{1}{\hbar}
G_\mu^{12}(\omega) e^{-i\eta\omega}
\right)\ ,
\label{epair1}
\end{eqnarray}
in keeping with the fact that the quantities entering the above equation
are the only ones depending explicitly on the anomalous density.
In the case of
$^{120}$Sn, $E^{\rm pho}_{\rm pair} = -3.9$ MeV, a number to be compared 
with $E_{\rm pair}{\rm (BCS+BH)} = -4.6$ MeV calculated making use
of the BCS relation 
\begin{equation}
E_{\rm pair}({\rm BCS+BH}) = 
- \sum_{\mu} 
\frac{2j_\mu+1}{2}
u_\mu v_\mu \Delta_{\mu}({\rm BCS+BH})\ ,
\label{epair2}
\end{equation}
%$$
%V_{\mu\mu^\prime} = 
%\langle (\mu^\prime m_{\mu^\prime})
%\overline{(\mu^\prime m_{\mu^\prime})}
%|V|
%(\mu m_{\mu})
%\overline{(\mu m_\mu)} \rangle_{\rm antisymmetrized} \ ,
%$$
where $\Delta_{\mu}({\rm BCS+BH})$ is the pairing gap calculated in the
BCS+BH approximation.

 We have also performed calculations using only 
two phonons, corresponding to 
the low-lying $\lambda^\pi = 2^+$ and $3^-$ modes ($n = 1$ in Table 1). 
The resulting average value of the pairing gap 
in the valence shell is 0.29 MeV, 
which represents about 50\% of the value (0.54 MeV) 
obtained using all the effective phonons listed in Table 1. 
It is seen that, while the low-lying phonons give the largest
contribution to the
induced pairing gap, the inclusion of the other modes  increases the gap
further.

 The perturbed single-particle levels 
in the valence shell in the calculation using only the low-lying 
$2^+$ and $3^-$ modes 
are less compressed around the Fermi level by only 100 keV in average 
as compared to those obtained in the full calculation, thus 
the perturbed spectrum is similar to Fig.\ 3(a). 

\section{Induced plus monopole interaction}

\begin{figure}
\begin{center}
% pcwalhalla:~/gs9/4-5pho/pho+p0/new
\epsfig{file=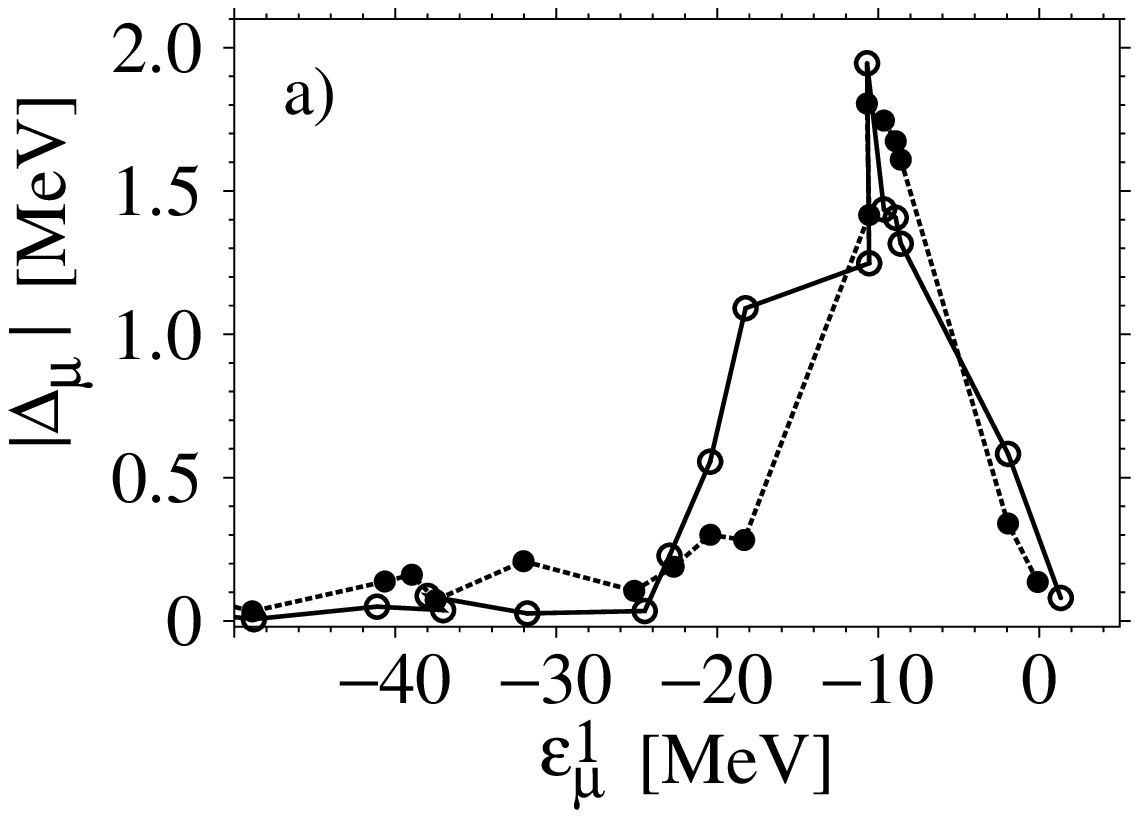,height=6cm}
\par\vspace{1em}
% pcwalhalla:~/gs9/4-5pho/pho+p0/new
\epsfig{file=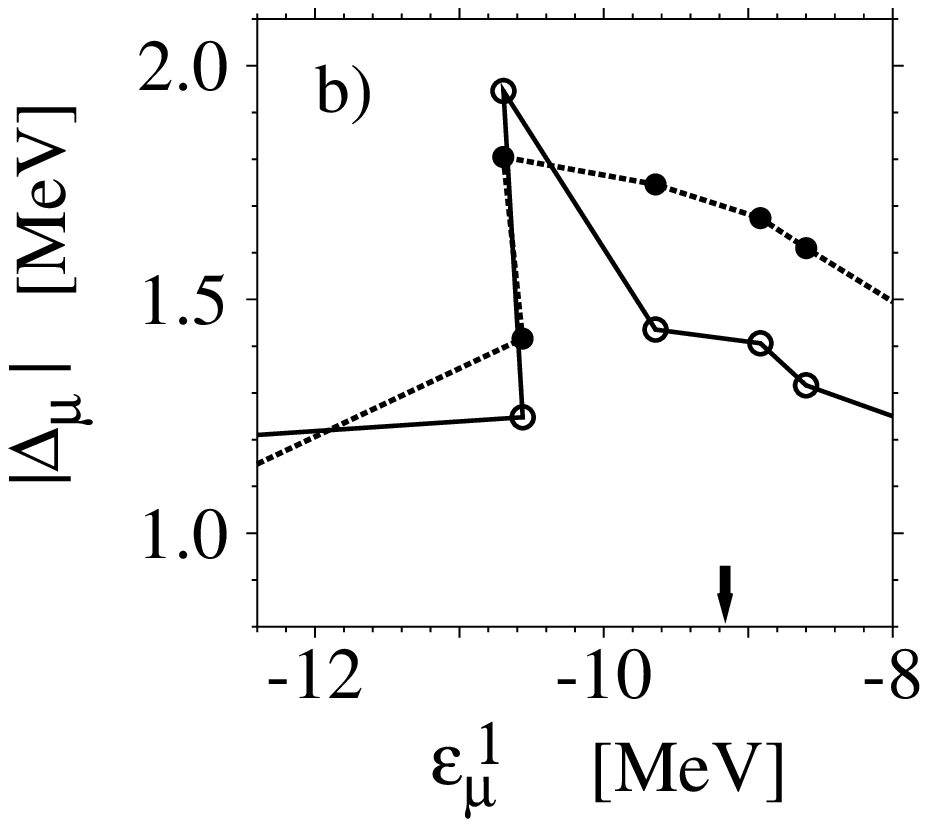,height=6cm}
\end{center}
\begin{center}
\parbox{12cm}{\small {\bf Figure 4.}
a) The pairing gaps obtained solving the Dyson equation 
including the monopole\-+induced interaction (open circles) are compared  
with the corresponding BCS+BH calculation (filled circles). 
b) A magnification of (a) near the Fermi level indicated by the arrow.
}
\end{center}
\end{figure}

The results presented in the previous section included only the contribution
of the phonon induced interaction. However, 
a realistic calculation of pairing correlations 
must include also a bare nucleon-nucleon interaction. We do this
in the present section, adding
to the self-energy of the phonon induced interaction used above (cf. Eq.\ (7))
the standard monopole (seniority) pairing field acting in the valence shell
(v.s.). 
Associated self-energy reads
\begin{eqnarray}
&&\hbar\Sigma^{12}_\mu{}^{P0} =
\hbar{\Sigma^{21}_\mu{}^{P0}}^\ast \nonumber\\
&&\hspace{1em}=
\left\{
\begin{array}{ll}
 -{\displaystyle \frac{g}{2}\sum_{\mu^\prime \in {\rm v.s.},\,m}
 (2j_{\mu^\prime}+1)\,R^{12}_{\mu^\prime m}(-\omega_{G+}^{\mu^\prime m})}\ , &
 \mbox{for $\mu$ in v.s.}\ , \\[1em]
 0\ , &
 \mbox{for other $\mu$}\ ,
\end{array}
\right.
\label{Sigma12P0}\\[1em]
&&
\hbar\Sigma^{11}_\mu{}^{P0} = \hbar\Sigma^{22}_\mu{}^{P0} = 0\ .
\label{Sigma11P0}
\end{eqnarray}
The experimental value of the pairing gap, deduced from the experimental 
odd-even mass difference, is $\Delta \simeq 1.4$ MeV
(cf. Chap.\ 2 in \cite{Bohr1}). 
The state-dependent gap obtained using $g= 20/A = 0.166$ MeV, plus the
induced phonon interaction used in the previous section, is shown in Fig.\ 4;
the average pairing gap in the valence shell 1.46 MeV is close to
the experimental value.
The strength $g =  0.166$ MeV should be compared with 
the value $g= 28/A = 0.233$ MeV needed to obtain the experimental value 
using the monopole-pairing field (\ref{Sigma12P0}) and (\ref{Sigma11P0}) 
without any phonon coupling (that is, performing a simple 
BCS calculation starting from the unperturbed single-particle 
spectrum with $m^* = 0.7m$). 

The enhancement of the pairing gap due to the induced phonon interaction
seems to be in contrast with the effect of the induced
interaction
in nuclear matter, where it usually decreases the pairing gap\footnote{
It was shown in \ ref.\ \cite{Sch96} 
that the sign of the polarization effect is density dependent 
in nuclear matter and that the effect reduces the gap at low densities. 
}. 
In fact, in neutron matter the induced interaction takes place mostly
through the exchange of $S=1$ excitations via the spin-spin part,
$(\mbox{\boldmath $s$}_1 \cdot \mbox{\boldmath $s$}_2)V_s$,
of the nucleon-nucleon force,
which leads to a repulsive interaction, and
tends to suppress the pairing gap.
In the case of finite nuclei the absence of collectivity associated to
the $S = 1$ modes should make  its contribution only marginal
with respect to the effect of the strongly collective low-lying surface modes
treated in this paper.

In order to get more insight in the result shown in Fig.\ 4, we decompose the 
anomalous self-energy $\Sigma_{\mu}^{12}$ in the sum of two terms, including
only the matrix elements of either the induced or the monopole interaction,  
\begin{equation}
 \hbar\Sigma^{12}_\mu(\omega_{G+}^{\mu m_0}) = 
\hbar\Sigma^{12}_\mu{}^{\rm pho}(\omega_{G+}^{\mu m_0}) +
 \hbar\Sigma^{12}_\mu{}^{P0}
 \label{sigma12}     \ .
\end{equation}
The imaginary parts of 
$\hbar\Sigma^{12}_\mu{}^{\rm pho}(\omega_{G+}^{\mu m_0}) $ and 
 $\hbar\Sigma^{12}_\mu{}^{P0}$ are negligible, and the real parts  have the 
same sign.
Also the state-dependent pairing gap, defined as 
\begin{equation}
\Delta_\mu =
\frac{\hbar\Sigma^{12}_\mu(\omega_{G+}^{\mu m_0})}
{Z_\mu(\omega_{G+}^{\mu m_0})}\ ,
\label{Deltamu}
\end{equation}
can then be separated into two contributions, 
\begin{equation}
 \Delta_\mu = \Delta^{\rm pho}_\mu + \Delta_\mu^{P0} \label{twocont}\ ,
\end{equation}
with
$$
\Delta^{P0}_\mu = \frac{\hbar{\Sigma^{12P0}_\mu}}
 {Z_\mu(\omega^{\mu m_0}_{G+})}\ .
$$
The monopole interaction has only a small effect on the values of
$Z_{\mu}$, since 
the interaction  does not act in the particle-hole channel:  
from Tables 2 and 3 it can be seen that the values of $Z_\mu$ 
obtained with or without the monopole interaction 
agree within 10\%
for all the valence orbitals.
From Table 3 it can also be seen that the  pairing gaps $\Delta_{\mu}$
receive similar contributions from $\Delta^{\rm pho}_\mu$ and
from $\Delta_\mu^{P0}$, except for the fragmented level $d_{5/2}$ where the
phonon gap is much larger.
In particular, the values of $\Delta^{\rm pho}_\mu$ are 
enhanced compared to
those obtained in section 3 without the monopole interaction, 
and listed in Table 2. 

\begin{table}
\begin{center}
\parbox{12cm}{\small {\bf Table 2.}
The result for the orbitals in the valence shell of the Dyson calculation 
with only the phonon-induced interaction including the effective coupling 
strengths. 
Only the real parts of the gaps are shown, 
since the imaginary parts are negligible. 
$\hbar\Sigma^{12}_\mu{}^{\rm pho}$ denotes  the value at the quasiparticle pole 
$\hbar\Sigma^{12}_\mu{}^{\rm pho}(\omega_{G+}^{\mu m_0})$.
}
\end{center}

\begin{center}
\small
\hspace*{10em}
\begin{tabular}{lrrrr}
\hline
\noalign{\rule{0ex}{1ex}}
& $\varepsilon^0_\mu$\hspace{1em} & $\Delta^{\rm pho}_\mu$ &
$\hbar\Sigma^{12}_\mu{}^{\rm pho}\hspace{-0.5ex}$ & 
$Z_\mu\hspace{0.5ex}$ \\
& [MeV] & [MeV] & [MeV] &  \\
\noalign{\rule{0ex}{1ex}}
\hline
\noalign{\rule{0ex}{1ex}}
h$_{11/2}$&$-$9.232   &  $-$0.642 & $-$1.049 & 1.63  \\
d$_{5/2}$ &$-$13.166  &  $-$0.951 & $-$2.396 & 2.52  \\
s$_{1/2}$ &$-$9.977   &  $-$0.519 & $-$0.866 & 1.67  \\
g$_{7/2}$ &$-$11.618  &  $-$0.183 & $-$0.231 & 1.26  \\
d$_{3/2}$ &$-$9.000   &  $-$0.353 & $-$0.578 & 1.64  \\
\noalign{\rule{0ex}{1ex}}
\hline
\end{tabular}
\end{center}
\end{table}

\begin{table}
\begin{center}
\parbox{12cm}{\small {\bf Table 3.}
The result of the Dyson calculation with 
both the phonon-induced interaction, including the effective coupling
strengths,  and the monopole (seniority) pairing force.
For the notations, see the text. 
Note that
$\hbar\Sigma_\mu^{12}{}^{\rm pho}$,
and 
$Z_\mu$ are 
abbreviations of
$\hbar\Sigma^{12}_\mu{}^{\rm pho}(\omega_{G+}^{\mu m_0})$, 
and 
$Z_\mu(\omega_{G+}^{\mu m_0})$, respectively.  
Again we show the real parts of the pairing gaps.
}
\end{center}

\begin{center}
\small
\hspace*{6em}
\begin{tabular}{lrrrrrr} \hline
\noalign{\rule{0ex}{1ex}}
{}&
 $\Delta_\mu\hspace{1ex}$&
 $\Delta^{\rm pho}_\mu$&
 $\hbar\Sigma^{12}_\mu{}^{\rm pho}\hspace{-1.0ex}$&
 $\Delta^{P0}_\mu$&
 $\hbar\Sigma^{12}_\mu{}^{P0}\hspace{-0.5ex}$&
 $Z_\mu$\hspace*{0.5ex}\\[0.5ex]
 & [MeV] & [MeV] & [MeV] & [MeV] & [MeV] \\[1.0ex]
\hline
\noalign{\rule{0ex}{1ex}}
h$_{11/2}$ & $-$1.400 & $-$0.868 & $-$1.287 & $-$0.531 & $-$0.788 & 1.48 \\
d$_{5/2}$  & $-$1.942 & $-$1.602 & $-$3.710 & $-$0.340 & $-$0.788 & 2.32 \\
s$_{1/2}$  & $-$1.429 & $-$0.919 & $-$1.419 & $-$0.510 & $-$0.788 & 1.54 \\
g$_{7/2}$  & $-$1.244 & $-$0.677 & $-$0.942 & $-$0.566 & $-$0.788 & 1.39 \\
d$_{3/2}$  & $-$1.311 & $-$0.792 & $-$1.204 & $-$0.518 & $-$0.788 & 1.52 \\
\noalign{\rule{0ex}{1ex}}
\hline
\end{tabular}
\end{center}
\end{table}

Finally, in Fig.\ 4 we compare the gaps obtained 
in the present calculations with
a BCS+BH calculation of the kind performed in [4], but including the same
monopole interaction and the same effective particle-vibration couplings as used
here. 
It is seen that the Dyson calculation leads to gaps which are
lower again by about 200 keV in average in the valence shell  
as compared to the BCS\-+BH results. 
This is because  
nucleons spend, in the Dyson scheme, part of their time
in more complicated configurations.

\section{Conclusions}

One can conclude,  that in the case studied the induced phonon 
interaction contributes substantially to pairing correlations: 
the strength of the monopole interaction, needed to 
obtain the experimental value of the pairing gap, 
is strongly reduced when the phonon contribution is included, 
as the gaps receive almost equal contributions from  
the monopole and from the induced interaction.

It is also seen that  a quantitative 
estimate of the pairing induced interaction requires  a self-consistent
treatment of the particle-vibration renormalization processes, as the
one performed here solving the Dyson equation. 
In fact 
the self-consistent
treatment lowers the gaps by about $20 \%$
as compared to the empirical treatment
of the effective mass processes, 
which does not  include a detailed description 
of the breaking of single-particle strength. 

%\vspace{12pt}

%\begin{center}
% osfmi:/temp/terasaki/bcs/lowest/dltcmp.ps
%  \epsfig{file=dltcmp.ps,height=5cm} 
%\end{center}

\vspace{12pt}

\end{document}